\documentstyle[preprint,aps,epsf]{revtex}

\draft
\tightenlines
\title{Relaxation at late stages in an entropy barrier model for\\
       glassy systems} 
\author{K. P. N. Murthy and K.W. Kehr}
\address{Institut f\"ur Festk\"orperforschung, Forschungszentrum
  J\"ulich GmbH,  D-52425 J\"ulich, Germany}
\begin{document}
\maketitle
\begin{abstract}
  The ground state dynamics of an entropy barrier model 
proposed recently for describing relaxation of glassy
systems, is considered. 
At late stages of evolution the
dynamics can be described by a simple variant of the Ehrenfest urn
model. Analytical expressions  
for the relaxation times from arbitrary initial 
states to the ground state are derived. Upper and lower bounds for the
   relaxation times as a function of system size are obtained. 
\end{abstract}
\pacs{  }

The Ehrenfest urn\cite{ehrenfest}  model has played a crucial 
role in formulating 
and clarifying several fundamental and subtle 
concepts of statistical mechanics. 
In this model distinguishable balls are placed in two boxes (or urns).
The dynamics consists of picking up a ball randomly and transferring it
from its box to the other. 
Concepts like what does  one mean by an equilibrium state, how does the system
approach, eventually reach and thereafter persist in equilibrium, 
and the meaning of fluctuations that
take the system away from its equilibrium state, become transparent 
when considering the time evolution in the Ehrenfest urn  
model. Interest in the Ehrenfest urn model, its variants 
and its generalizations 
has been revived recently following the work of Ritort
\cite{ritort};  see \cite{french1,french2}, for some 
subsequent work on Ritort's and related models. 

Ritort's model essentially 
describes the relaxation of a nonequilibrium system. The ground state
dynamics of Ritort's model at late stages of evolution is describable
by a simple variant of the classical Ehrenfest urn
model{\cite{lipowski}.
In this letter we employ  first passage time formulation and obtain
analytical expressions for the relaxation times from an arbitrary
initial state
of (the variant of) the Ehrenfest urn model. 
 
 In the model proposed by Ritort, $N$ distinguishable 
balls are placed in $N$ boxes. Energy is defined as minus the number of empty
boxes. The dynamics is defined as follows. 
Select a ball and a box randomly and independently. If the selected box 
is non-empty deposit the ball in the box. If the box is empty, the 
transfer of the ball  would increase the energy; hence carry out the
transfer only with  
a probability given by the Boltzmann  factor $\exp[-\Delta E / K_B T]$, 
where $\Delta E\ (=1)$  
is the increase in the 
energy, $K_B$ is the Boltzmann constant and $T$ 
is the temperature. The static properties 
of this model can be calculated exactly. 
For example at the ground state, which is $N$-
fold degenerate, the energy per particle is $-1+1/N$. 
All the balls are in one 
of the $N$ boxes at the ground state. Ritort's model 
could explain several important 
characteristics of the glassy systems, like anomalously 
slow relaxation, aging, and hysteresis.
Ritort's model is perhaps the first  model in which only 
entropy barriers are 
present explicitly. The ground state dynamics  of the model is equivalent to an 
$N$ -  urn generalization and variant of the Ehrenfest model. Moving a 
ball to an empty box is  
disallowed. Thus once a box becomes empty it
stays empty for ever.
As a result the number of empty boxes increases with time. 
The energy decreases with time and eventually
reaches its ground state value. The relaxation of the energy 
is slow and becomes slower with
time. This is easy to understand at least qualitatively, since lowering the occupancy 
of a box becomes less and less likely as the number of balls in the box becomes fewer and
fewer. Let us denote by $\Omega(N, N)$ the set of all possible configurations of distributing 
$N$ balls in $N$ boxes. Let $\tau$ denote the time it takes for the system to relax to the 
ground state, averaged over the ensemble $\Omega(N, N)$ of initial configurations. It 
is clear that the system before reaching the ground state, arrives invariably at one 
of the simpler non equilibrium configurations with only two boxes non empty. The 
other $N-2$ boxes  have since been emptied during the evolution.  Let us denote by $\Omega(2,N)$
the set of all possible states with $N$ balls in $2$ boxes. Also let
$\tau _1$ denote the relaxation time  
averaged  over the ensemble $\Omega(2,N)$ of initial 
states.  In a recent work, Lipowski\cite{lipowski} showed  
that $\tau _1 $ is less than but almost equal to $\tau $, for large
$N$.

The purpose of this letter can be stated as follows. Lipowski\cite{lipowski}
derived 
analytical expression for the relaxation 
from one of the simple states belonging to the set $\Omega(2,N)$. This state 
consists of only one ball in one box and the remaining $N-1$ balls in the other. Relaxation 
times from other states like $k$ balls in one box and the rest in the other, were calculated 
through a set recursion relations. Lipowski's work is  
a clever application of the formulation of Darling and
Siegert\cite{darling}, for first passage times of random walks.
In this letter we employ a simpler and more recent first passage time
formulation\cite{murthy}. This method  enables  us to obtain 
closed-form  expression for the relaxation 
time from an arbitrary initial state, a task  
which has not been possible in Lipowski's formulation. 
More importantly, employing the analytical expressions,  we establish useful upper
and lower bounds for the relaxation times from various initial states in terms of the system
size.

We first consider the case with even number of balls distributed in two boxes. Accordingly 
let $2N$ be the total number of balls in the system. Let us consider a configuration with 
$k$ balls in one box and the remaining $2N-k$ balls in the other. We say the system is in state 
$k$ if the number of balls in the box of lower occupancy equals
$k$. Thus $k$ is less than or equal to $N$. Let 
$\hat{G}_{k,k-1}(\nu) $ denote the probability for the system to go from the state $k$ to the 
state $k-1$, for the {\it first time} in exactly $\nu$ time
steps. Thus $\nu$ is a {\it discrete} random variable 
and is called the First Passage Time(FPT). The first passage from $k$ to $k-1$ can happen 
 in two ways: (a) Select a ball from the box containing $k$ balls and transfer it to the other 
taking one time step. The probability for this is $k/2N$. At the end of this step we are 
in the target state $k-1$. Hence $\nu$, the FPT is unity. (b) Select a ball from the box containing
$2N-k$ balls and move it to the other box taking one step. The probability for this 
is $(2N-k)/2N$. At the end of this step we are in the state $k+1$. Now
make a first passage 
from the state $k+1$ to the state $k-1$ in the remaining $\nu-1$
steps. The above considerations  
hold good for all $k\le N-1$. For $k=N$, we find that since both the boxes contain $N$ balls 
each, selecting a ball from either and moving it to the other leads to 
the target state $k-1$, and the FPT is unity. The state $N$ is thus
reflecting. We have therefore,
\begin{eqnarray}\label{fpteven}
\hat{G}_{k,k-1}(\nu)
&=&{{k}\over{2N}}\delta_{\nu,1}+{{2N-k}\over{2N}}\hat{G}_{k+1,k-1}(\nu-1), \quad\quad
                                             \rm{for}\quad k=1,2,\cdots ,N-1,\nonumber\\
\hat{G}_{N,N-1}(\nu)&=&\delta _{\nu ,1}.
\end{eqnarray}
The above is the complete set of $N$ equations for the FPT
densities. Since the system starting from state $k+1$ can not reach
the state $k-1$ without visiting the state $k$,
$\hat{G}_{k+1,k-1}(\nu)$ in the above can be expressed as a
convolution 
given by,
\begin{equation}
  \label{convolution}
  \hat{G}_{k+1,k-1}(\nu)=\sum_{\eta =1}^{\nu -1}\hat{G}_{k+1,k}(\eta)\hat{G}_{k,k-1}(\nu-\eta).
\end{equation}
The convolution in the above holds good in general for one dimensional
problems with nearest neighbour hopping. 

When the number of balls is odd, say $2N-1$, the equations for the FPT
densities remain formally the same except that, now the state $N-1$ is
{\it reflecting}, in the following sense. Consider the first passage
from the state $N-1$ to the state $N-2$, in $\nu$ steps. There are two
ways: (a) Move a ball from the box containing $N-1$ balls to the other
box, in one step;  the probability for this is $(N-1)/( 2N-1)$. At the
end of this step we are in the target state $N-2$. Hence the FPT is
unity. (b) Move a ball
from the box containing $N$ balls to the other box, 
taking one time 
step. The probability for this is $N/(2N-1)$. At the end of this step we
are in state $N-1$, the same state we started with. Now make a first passage from
this state $N-1$ to the target state $N-2$ in the remaining $\nu-1$
steps. Thus for odd number of balls we get,
\begin{eqnarray}\label{fptodd}
\hat{G}_{k,k-1}(\nu)&=&{{k}\over{2N-1}}\delta _{\nu,1} +
{{2N-1-k}\over{2N-1}}\hat{G}_{k+1,k-1}(\nu-1), \quad\quad\rm{for}\quad
 k=1,2,\cdots,N-2\nonumber \\
\hat{G}_{N-1,N-2}(\nu)&=&{{N-1}\over{2N-1}}\delta_{\nu,1}+{{N}\over{2N-1}}\hat{G}_{N-1,N-2}(\nu-1).
\end{eqnarray}
The above constitute the complete set of $N-1$ equations for the FPT
densities for the case with odd number of
balls. To solve the recursion relations (\ref{fpteven}) 
and (\ref{fptodd})  for the first passage time densities  
we employ generating function technique. 

Let $G_{i,j}(Z)$ denote the generating function for the FPT, defined
as,  
\begin{equation}
  \label{gf}
  G_{i,j}(Z)=\sum_{\nu=1}^{\infty}Z^{\nu} \hat{G}_{i,j}(\nu).
\end{equation}
Multiplying both sides of Eqns.(\ref{fpteven}) by $Z^{\nu}$ and summing
over $\nu$ from $1$ to $\infty$, we get,
\begin{eqnarray}
  \label{fptgf}
  G_{k,k-1}(Z)&=&Z{{k}\over{2N}} + Z{{2N-k}\over{2N}}G_{k+1,k-1}(Z), \quad\quad
  \rm{for}\quad  k=1, \cdots , N-1\nonumber\\
G_{N,N-1}(Z)&=& Z.
\end{eqnarray}
 
Equivalent relations, not given here, can be obtained for the case
with odd number balls. 
From Eqns.(\ref{fptgf}), a terminating continued fraction relation
for $G_{k,k-1}(Z)$ can be derived by noting that $G_{k+1,k-1}(Z)
=G_{k+1,k}(Z)\times G_{k,k-1}(Z)$, by virtue of convolution theorem.
Substituting the convolution in
Eqns. (\ref{fptgf}), we get,
\begin{eqnarray}
  \label{cf}
  G_{k,k-1}(Z)&=&{{ Z{{k}\over{2N}} }\over{1- Z{{2N-k}\over{2N}}G_{k+1,k}
  (Z) }}, \quad\quad \rm{for}\quad  k=1,\cdots,N-1\nonumber\\
  G_{N,N-1}(Z) &=&Z.  
\end{eqnarray}
In fact, by convolution we have $G_{m,0}(Z)=\prod_{k=1}^{m}G_{k,k-1}(Z)$,
for $m=1,\cdots,N$. Thus in principle we have obtained the
distribution of relaxation
time from an arbitrary state belonging to $\Omega(2,N)$, to the
zero ground state, though the expressions are in $Z$ space.

For calculating the mean first passage time(MFPT), from the state $k$ to
the state $k-1$, we differentiate $G_{k,k-1}(Z)$ with respect to $Z$ and
set $Z=1$. Let $F_{k,k-1}$ denote the MFPT from $k$ to $k-1$. 
For the  problem with even number of balls, we get,
\begin{eqnarray}
  \label{mfpt}
 F_{k,k-1}&=&{{2N-k}\over{k}}F_{k+1,k}+{{2N}\over{k}}, \quad\quad\rm{for}\quad k=1,\cdots ,N-1\nonumber\\
F_{N,N-1}&=&\left( {{1}\over{2}}\right)\  {{2N}\over{N}}. 
\end{eqnarray}
The above can be cast in a convenient matrix notation,
\begin{equation}
  \label{fafpu}
  |F\rangle =A|F\rangle +|U\rangle ,
\end{equation}
where $|F\rangle $ is a column vector $(F_{1,0}\quad F_{2,1}\quad \cdots ,
F_{N,N-1})^{\dagger}$ and $|U\rangle $ is the column vector representing the
inhomogeneities, $(2N/1\quad  2N/2\quad  \cdots\quad  1)^{\dagger}$. Here the
superscript ${}^{\dagger}$ denotes the transpose operation. The
$N\times N$ matrix $A$ has
elements given by $A_{i,j}=\delta _{i,j-1}\times (2N-i)/i$.   We
can
cast Eq. (\ref{fafpu}) as $B|F\rangle =|U\rangle $ where $B=I-A$. The
matrix $B$ has 
all its diagonal elements unity; all the other elements except those
in the first upper diagonal are zero. The matrix elements of $B$ are
given by,
\begin{equation}
  \label{bmatrix}
  B_{i,j}=-\left( {{2N-i}\over{i}}\right) \times \delta _{i,j-1} +  \delta _{i,j}.
\end{equation}
We give below the matrix $B$ explicitly, to facilitate easy
visualization of solutions we are going to derive shortly.  
\begin{equation}
  \label{matrixb}
  B=\left( \begin{array}{cccccc}
          \quad 1\quad & -\left( {{2N -1}\over{1}}\right)
 &\quad 0 \quad  &\quad 0 \quad &\quad 0 \quad  &\quad \cdot\quad  \\
           0 & 1& -\left( {{2N-2}\over{2}}\right) &0&\cdot&\cdot    \\
         \cdot &\cdot&\cdot&\cdot&\cdot&\cdot        \\
 \cdot &\cdot  &\cdot&\cdot&\cdot&\cdot        \\
          0&0&0&0&1&-\left( {{2N-\left(N-1\right) }\over{2N}}\right)\\
          0&0&0&0&0&1
           \end{array}\right).
\end{equation}
To calculate the $m-$th  element $F_{m,m-1}$ of the vector $|F\rangle$,
we replace the $m$-th column of the matrix $B$ by the vector
$|U\rangle$. Let the matrix thus formed be denoted by
$B_{(m)}$. Then $F_{m,m-1}$ is given by Cramer's rule,
\begin{equation}
  \label{kramer}
  F_{m,m-1}= {{ D[B_{(m)}]}\over{D[B]}}, 
\end{equation}
where $D[\cdot ]$ denotes the determinant. First we observe that the
determinant of the matrix $B$ is unity. The problem reduces to
calculating the determinant of $B_{(m)}$. 

Let us consider the case
with $m=1$. The determinant of $B_{(1)}$ can be easily 
written down by inspection as,  
 \begin{eqnarray}
   \label{dofb1}
   F_{1,0} &= & D\left[ B_{(1)} \right] = {{2N}\over{1}} +
                       {{2N}\over{2}} \left( {{2N-1}\over{1}} \right)+
                       {{2N}\over{3}} \left(  {{2N-1}\over{1}} \right)
                       \left( {{2N-2}\over{2}} \right) \nonumber\\ 
\quad &\quad&         +\cdots\cdots\nonumber\\  
\quad &\quad&         +{{1} \over{2}}\quad  {{2N}\over{N}} \left(
                       {{2N-1}\over{1}} \right)
                       \left(  {{2N-2}\over{2}} \right) \cdots\left(
                        {{2N-(N-1)}\over{N-1}} \right),
\end{eqnarray}
which can be cast
 in a compact form as sum over products given by,
\begin{eqnarray}
  \label{f1zero}
  F_{1,0} & = & \sum_{n=1}^{N} {{2N}\over{n}} \left(
          1-{{1}\over{2}}\delta _{n,N} \right)
          \prod_{k=1}^{n-1} {{2N-k}\over{k}}.\nonumber\\
\quad    & = & \sum_{n=1}^{N-1} {}^{2N} C_{n} +\left(
          {{1}\over{2}}\right)  {}^{2N} C_N ,
\end{eqnarray}
where ${}^{(.)} C_{(.)}$ are the usual binomial coefficients. Noting that
${}^{2N}C_{n}$ is the same as ${}^{2N}C_{2N-n}$, we see immediately,
that
\begin{equation}
  \label{2fonezero}
  2\times F_{1,0}=\sum_{n=1}^{2N-1}{}^{2N}C_{n}.
\end{equation}
Add the binomial coefficients ${}^{2N}C_{0}=1$ and ${}^{2N}C_{2N}=1$ to
both sides of the equation above. We find that the right hand side
becomes $2^{2N}$, and we get,
\begin{equation}
  \label{adam}
  F_{1,0}=2^{2N-1}-1,
\end{equation}
which is precisely the expression derived by
Lipowski\cite{lipowski}. Note that in Lipowski's paper $N$ denotes the
total number of balls whereas here the total number of balls is taken
as even ($2N$), see eq. (\ref{adam})  or odd ($2N-1$), see
Eq. (\ref{fkzeroodd}). We have considered the two cases separately
for bringing out clearly the subtle difference in the reflecting boundary
while deriving the master equations, see Eq. (\ref{fpteven})
and Eq. (\ref{fptodd}). 

Let us now derive closed-form expressions for $F_{m,m-1}.$ To this 
end, we replace the $m$-th
column of the matrix $B$ by the vector $|U\rangle$ and construct the
matrix $B_{(m)}$, whose determinant gives,
\begin{equation}
  \label{fmminusone}
  F_{m,m-1}=\sum_{n=m}^{N-1}\left( {{2N}\over{n}}\right)  \prod_{k=m}^{n-1}{{2N-k}\over{k}}+ 
 \left( {{1}\over{2}}\right) \left( {{2N}\over{N}}\right)  \prod_{k=m}^{N-1}{{2N-k}\over{k}}.
\end{equation}
First we multiply both sides of the above equation by
$\prod_{k=1}^{m-1}(2N-k)/k\equiv (m/2N)\ {}^{2N}C_{m}$, and  get,
\begin{eqnarray}
  \label{eqnint}
  \left( {{m}\over{2N}}\right) {}^{2N}C_{m}\ F_{m,m-1}&=&\sum_{n=m}^{N-1}\left( 
            {{2N}\over{n}}\right)  \prod_{k=1}^{n-1}{{2N-k}\over{k}}+ 
            \left( {{1}\over{2}}\right) \left( {{2N}\over{N}}\right)  
            \prod_{k=1}^{N-1}{{2N-k}\over{k}}.\nonumber\\
    \quad &=&\sum^{N-1}_{n=m}{}^{2N}C_{n}+\left({{1}\over{2}}\right){}^{2N}C_{N}. 
\end{eqnarray}
Now add to both sides of the above equation the term
$\sum_{n=1}^{m-1}{}^{2N}C_{n}$, and get,
\begin{equation}
  \label{inteq2}
 \sum_{n=1}^{m-1}{}^{2N}C_{n}+\left( {{m}\over{2N}}\right)
 {}^{2N}C_{m} F_{m,m-1}=\sum^{N-1}_{n=1}{}^{2N}C_{n}  +\left(
 {{1}\over{2}}\right) {}^{2N}C_{N}. 
\end{equation}
We see immediately that
\begin{equation}
  \label{inteq3}
  2\times\left[ \sum_{n=1}^{m-1}{}^{2N}C_{n}+\left( {{m}\over{2N}}\right)
 {}^{2N}C_{m} F_{m,m-1}\right] 
=\sum^{2N-1}_{n=1}{}^{2N}C_{n}. 
\end{equation}
If we add now ${}^{2N}C_{0}=1$ and ${}^{2N}C_{2N}=1$, to both sides of the
above equation, we find the right hand side is simply $2^{2N}$. We get  
\begin{equation}
  \label{finalfm}
  F_{m,m-1}=\left( {{2N}\over{m}}\right) \left( {{1}\over{
  {}^{2N}C_{m} }} \right) \left[ \left(
  2^{2N-1}-1\right)
  - \sum_{n=1}^{m-1} {}^{2N}C_{n}\right].
\end{equation}
It is easily seen that if we substitute $m=1$ in the above we recover
Lipowski's result\cite{lipowski}, also derived explicitly in this
paper, see Eq.(\ref{adam} ). The relaxation time from any state
$k$ to the zero ground state can be obtained by summing
Eq. (\ref{finalfm}) over $m$ from $1$ to $k$. Thus we get,
\begin{equation}
  \label{fkzero}
  F_{k,0}=\sum_{m=1}^{k}\left( {{2N}\over{m}}\right) \left( {{1}\over{
  {}^{2N}C_{m} }}\right) \left[ \left(
  2^{2N-1}-1\right)
  - \sum_{n=1}^{m-1} {}^{2N}C_{n}\right].
\end{equation}
For odd number of balls the derivation proceeds in the same way, and
we get
\begin{equation}
  \label{fkzeroodd}
  F_{k,0}=\sum_{m=1}^{k}\left( {{2N-1}\over{m}}\right) \left( {{1}\over{
  {}^{2N-1}C_{m} }}\right) \left[ \left(
  2^{2N-2}-1\right)
  - \sum_{n=1}^{m-1} {}^{2N-1}C_{n}\right].
\end{equation}

Now that we have analytical expression for the relaxation time, $F_{k,0}$ from an
arbitrary state, we can estimate how much it deviates from 
$F_{1,0}$, when the system size goes to infinity. From
Eq. (\ref{finalfm})       
it is clear that 
\begin{equation}
  \label{hierarchy}
  F_{1,0}\  >\ F_{2,1} \ >\ F_{3,2}\ >\ \cdots\ >\ F_{N-1,N-2}\ >\ 
  F_{N,N-1}\ \left( =\ 1\right) .
\end{equation}
In fact for large $N$, we have, from Eq. (\ref{finalfm}),
\begin{equation}
  \label{largeN1}
  F_{m,m-1}\ \ \ {}^{\ \ \sim\ \ }_{N\to\infty}\ \ \ F_{m-1,m-2}\
  \left( {{m-1}\over{2N}}\right),
\end{equation}
which implies that, 
\begin{equation}
  \label{asymp1}
  F_{m,m-1}\ \ \ {}^{\ \ \sim\ \ }_{N\to\infty}\ \ \ F_{1,0} \ {{ \left( m-1\right) !}\over{
  \left( 2N\right) ^{m-1} }}.
\end{equation}
Since,  $F_{k,0}=F_{1,0}+F_{2,1}+\cdots +F_{k,k-1}$, we have,
\begin{equation}
  \label{asymp2}
  F_{k,0}\ \ \ {}^{\ \ \sim\ \ }_{N\to\infty}\ \ \ F_{1,0}\left[
    1+{{1}\over{2N}}+{{2}\over{(2N)^2}}+{{6}\over{(2N)^3}}+
    \cdots+{{ (k-1)!}\over{ (2N)^{k-1} }}\right],
\end{equation}
for all $k$. Thus to the order of $N^{-1}$,  we have  
\begin{equation}
  \label{asymp3}
  F_{k,0}\ \ \ {}^{\ \ \sim\ \ }_{N\to\infty}\ \ \ F_{1,0}\left( 1+{{1}\over{2N}}\right),
\end{equation}
for all $k\ge 2$ and the correction is independent of $k$. In fact it is
easily shown from Eq. (\ref{asymp2}) that,
\begin{equation}
\label{bounds}
F_{1,0}\left( 1+{{1}\over{2N}}\right) \ <\ F_{k,0}\ < F_{1,0}\left( 1+{{1}\over{N}}\right)
\end{equation}
for all $k\ge 2$, when $N$ is large.
Thus it is clear
that, indeed  $F_{1,0}$ is the principal  time scale in
the problem and relaxation time  from any other state is only negligibly
greater than $F_{1,0}$ for large systems.   

This letter, in a way complements the work
of Lipowski\cite{lipowski}. We have obtained
analytical expressions for the relaxation times to the ground state
(with all the balls in one box and the other box empty),  starting from an
arbitrary initial state (with say, $k$ balls in one box
and the rest in the other). We have shown that the relaxation time
from the simple state $k=1$ (with one ball in one box 
and the rest in the other)
sets the principal time scale in the problem; relaxation from other
states takes negligibly more time than this, for large systems. 
A natural question that arises 
in this context is whether we can construct a simple 
two-urn analogue of Ritort's model. To this end we need to
suitably modify the energy function defined over the states of
$\Omega (2,N)$. For example we can define energy as minus the absoluted
value of the difference of the number of balls in the two boxes. 
{\it i.e.}, $E(n)=2n-N$, where $n\, \le \, N/2$ is the number of balls in the
lower occupancy box(defining the state of the system) and $N$ is the
total number of balls. It is easily seen  that the state $n=0$ is the
minimum energy ground state with $E(0)=-N$. The maximum energy state
is $n=N/2$, with $E(N/2)=0$. It can be shown that 
an  arbitray state $k$ relaxes 
to the ground state in a time 
given by $N\sum_{m=1}^{k}m^{-1}$, at zero temperature. However as the 
temperature increases, the relaxation of energy to its equilibrium
value (at that temperature) becomes faster.
It is indeed  worthwhile 
investigating if other features like 
hysteresis
 and aging are also present in this
simple model.

KPNM thanks Forschungszentrum J\"ulich, for the hospitality extended
to him at the Institut f\"ur Festk\"orperforschung.


\begin{thebibliography}{xxx}
\bibitem{ehrenfest}
  Ehrenfest P 1912 {\it The conceptual foundations of the statistical 
  approach in mechanics}, translated by M J Moravcsik (Cornell
  University Press, 
  Ithaca, New York) 1959; see also Klein M J ed. 1959 {\it Paul
  Ehrenfest, collected scientific papers} (North-Holland, Amsterdam)
\bibitem{ritort}
  Ritort F 1995 {\it Phys. Rev. Lett.} {\bf 75} 1190 
\bibitem{french1}
  Godreche C, Bouchaud J P and Mezard M 1995 {\it J. Phys. A:
  Math. Gen. } {\bf 28} L603
\bibitem{french2}
  Godreche C and Luck J M 1996 {\it J. Phys. A: Math. Gen.} {\bf 29} 1915
\bibitem{lipowski}
  Lipowski A 1997 {\it  J. Phys. A: Math. Gen. } {\bf 30} L91
\bibitem{darling} Darling D A and Siegert A F J 1953 {\it
  Ann. Math. Stat.}  {\bf 24} 624
\bibitem{murthy} Murthy K P N and Kehr K W 1989 Phys. Rev A{\bf 40}
  2082; see also Zwerger W and Kehr K W 1980 Z. Phys B{\bf 40} 157. 
\end{thebibliography}
\end{document}